\documentstyle[12pt,epsfig]{article}

\begin{document}
\title{Defect fugacity, Spinwave Stiffness and $T_c$ 
of the 2-d Planar Rotor Model} 

\author{Surajit Sengupta$^1$\thanks{{\it on leave from}: Material Science 
Division, Indira Gandhi Centre for Atomic Research, Kalpakkam 603102, India},
Peter Nielaba$^2$, and K. Binder$^1$}

\date{\today}

\maketitle

\begin{center}
$^1$ Institut f\"ur Physik, \\
Johannes Gutenberg Universit\"at Mainz, \\
55099, Mainz, Germany 

\vskip .5cm 

$^2$ Universit\"at Konstanz,  \\ Fakult\"at f\"ur 
Physik, Fach M 691, \\  
78457, Konstanz, Germany
\end{center}

\begin{abstract}
 We obtain precise values for the fugacities of vortices 
 in the 2-d planar rotor model from 
 Monte Carlo simulations in the sector with {\em no} vortices. The
 bare spinwave stiffness is also calculated  
 and shown to have significant anharmonicity. Using these as inputs in 
 the KT recursion relations, we predict the temperature $T_c = 0.925$, 
 using linearised equations, and $T_c = 0.899 \pm .005$ using next higher 
 order corrections, at which vortex unbinding commences in the unconstrained 
 system. The latter value, being in excellent agreement with all recent 
 determinations of $T_c$, demonstrates that our method 1) constitutes a 
 stringent measure of the relevance of higher order terms in KT theory 
 and 2) can be used to obtain transition temperatures in similar systems 
 with modest computational effort. 
\end{abstract}
\newpage

\noindent
\underline{{\bf Introduction:}}\,\, The phase behaviour of isotropic magnets
and related systems in two dimensions is a particular challenge since famous
theorems exclude long-range order\cite{MW}, but nevertheless phase transitions
occur in models such as the two- dimensional XY ferromagnet\cite{SK} or the 
planar rotor model ($S^x_i = \cos\phi_i, S^y_i = \sin\phi_i$, where $\bf S_i$
is a two component spin at the site $i$ with unit magnitude and orientation
$ 0 \leq \phi_i < 2 \pi$ ),
\begin{equation}
\beta {\cal H} = -\frac{1}{T}\sum_{<ij>}\cos(\phi_i-\phi_j),
\end{equation} 
the sum extends once over all nearest neighbor pairs of the (square) lattice, 
and $T$ is the reduced temperature (the Boltzmann constant $k_B = \beta^{-1}T$
is taken to be $1$ throughout). Originally\cite{SK} it was proposed that 
a critical temperature $T_c$ occurs where the correlation length $\xi$ 
describing the decay of the correlation function $g(\bf r) = <\bf S(0) \bf S(\bf r)>$ with distance $\bf r$, and the susceptiblity $\chi = \sum_{\bf r}
<{\bf S}(0)\cdot{\bf S}({\bf r})>/T$ diverge according to power laws,
\begin{equation}
\xi \sim t^{-\nu}, \chi \sim t^{-\gamma}, t \equiv T/T_c - 1
\end{equation} 
$\nu, \gamma$ being the usual critical exponents. However, Kosterlitz and 
Thouless (KT)\cite{KT1,KT2} developed a completely different scenario, based on 
the unbinding of vortex- antivortex pairs, yielding an essential singularity,
\begin{eqnarray}
\ln \xi & = & \ln \xi_0 + b t^{-\bar \nu}, \\ \nonumber
\chi    & \propto & \xi^{2-\eta}
\end{eqnarray}
\noindent
where $\xi_0, b$ are nonuniversal constants, while an approximate 
renormalization group treatment\cite{KT1,KT2} predicted that the exponents 
$\bar \nu, \eta$ take the universal values,
\begin{eqnarray}
\bar \nu & = & 1/2, \\ \nonumber
\eta & = & 1/4.
\end{eqnarray}

For $T<T_c$, spin-wave theory remains essentially valid and $g(r) 
\sim r^{-\eta(T)}$ where $\eta(T)$ increases smoothly with increasing 
temperature from $\eta(T=0) = 0$ upto $\eta(T_c)\equiv \eta = 1/4$. The
spinwave stiffness $K(T)$ (in which we have absorbed a factor of $1/T$ as in 
Eq.(1)) smoothly decreases and also involves a universal 
ratio at $T_c, K(T_c) = 2/\pi$\cite{KT2,NK}. A related critical behaviour is 
predicted for the superfluid- normal fluid transition of helium in 
two- dimensions\cite{NK}, for the roughening transition of interfaces\cite
{chui}, transitions of adsorbed layers on surfaces to modulated structures
incommensurate with the substrate periodicity, etc. Thus, this problem has
found widespread interest\cite{revw}. A particularly interesting ---- but also 
still controversial---- extension deals with two- dimensional melting
\cite{HN,KJS,wmb}.

However, both the physical mechanism for the vortex- antivortex pair 
dissociation at $T_c$ and the resulting predictions have been questioned 
many times (eg.\cite{zscal,patra}). The theory\cite{KT1,KT2,NK,joska,ohta,
naka,cardy,amit,kadz}
involves problematic assumptions such as the decoupling of vortex and 
spinwave excitations; and numerical analyses\cite{patra,SH,
TC,HC,GB,JN,OL,JKK,Janke} often are not
fully convincing although usually the KT theory is favoured. Monte Carlo 
studies are
difficult since $\xi$ increases so strongly as $t$ gets small (eg. $\xi > 40$
lattice spacings for $t\leq.1$), and so it is questionable whether the 
asymptotic critical region is reached. Even studies for very large lattices 
($1200\times1200$) still reveal problems with Eq.(4)\cite{JN}, and the 
simulation data can well be fitted to Eq.(2) if (albeit rather large) 
corrections to scaling are taken into account\cite{JN}. The most recent 
analyses, in fact, point towards the need of considering logarithmic
corrections\cite{JKK,Janke}.

In the present paper we hence follow a different strategy for a Monte Carlo
test of the KT renormalization approach, avoiding the brute force methods of
Refs.\cite{patra,SH,TC,HC,GB,JN,OL,JKK,Janke}. Namely we test the KT scenario
 by obtaining the proper
input parameters for the renormalization group flow equations, which then are
solved numerically. In this way a stringent consistency test is possible which
is different from all previous approaches to the problem.  
\vskip 1cm 

\noindent
\underline{{\bf Monte Carlo estimation of input parameters to the KT theory}}\,
The KT theory\cite{KT1,KT2,NK,chui,revw} can be cast in the framework of a 
two parameter
renormalization flow for the spinwave stiffness $K(l)$ and the fugacity of
vortices $y(l)$, where $l$ is related to the considered length scale as $l = 
\ln(r/a)$, where $a$ is the lattice spacing. These flow equations in terms 
of the scaled variables $x = (2-\pi K)$ and $y' = 4 \pi y$ and upto 
next to leading order\cite{amit} are,
\begin{eqnarray}
\frac{d x}{d l} & = & y'^2 - y'^2 x, \\ \nonumber
\frac{d y'}{d l} & = & x y' + \frac{5}{4} y'^3.
\end{eqnarray}
Using a linearised version of these equations (i.e. keeping only the first 
terms on the right hand side of these equations) and using the approximate 
initial conditions $y(l=0) \simeq \exp(-10.2/2 T)$ and $K(l=0) = 1/T$ --
a result from harmonic spin wave theory strictly valid at $T \to 0$, 
Kosterlitz\cite{KT2} found that a non- trivial fixed point $K(l=\infty) = 
2/\pi, y(l=\infty) = 0)$ exists (cf. Eq. (3) above) but the resulting estimate 
for $T_c \simeq 1.35$ is rather different from the current best estimates 
$T_c = 0.895\pm .005\,$\cite{GB,OL,JKK}. Does this discrepancy mean that the 
KT scenario does not work?

Such a conclusion would clearly be premature, however, because the above 
assumption implies that even at $T=T_c$ one can still take the unrenormalized
zero temperature value of the spin wave stiffness as a starting value for 
the recursion, Eq.(5). To test this assumption we have obtained $K$ (and $y$) 
from Monte Carlo simulations. Two sets of simulations are carried out. The 
first set uses the full Hamiltonian, Eq.(1), while the second set uses the 
constraint that neither vortices nor antivortices can form\cite{benav}.
Note that an elementary plaquette of the square lattice contains a vortex
or an antivortex, if the angles $\phi_i$ of the spins $1,2,3,4$ at the 
corners of the plaquette (labelled anticlockwise) satisfy the condition
$\sum_{i=1}^4 \Delta \phi_i = \pm 2\pi$, where we have defined $\Delta \phi_i =
\phi_{i+1}-\phi_i, \phi_5=\phi_1$. If there are only spin wave excitations 
in the system,$\sum_{i=1}^4 \Delta \phi_i = 0$ for all plaquettes. Hence the
no- vorticity constraint in the Monte Carlo elementary step (which involves an 
attempt to replace $\phi_i$ by a randomly chosen $\phi_i'$, with $0\leq 
\phi_i' < 2 \pi$) considers whether $\sum_{i=1}^4 \Delta \phi_i = 0$ is 
still true for
this trial configuration for all the four adjoining plaquettes to which the 
site $i$ belongs. If the constraint is not true, the trial move is 
automatically rejected. Note that we always start the simulation from a vortex 
free initial fully aligned state ($\cos\phi_i=1$ for all $i$).

Fig.1 gives a plot of the inverse stiffness constant $K^{-1} = 4 \pi 
\ln<M^2>/\ln N$, where $<M^2> = T\chi_N$. Note that while $K^{-1}$ diverges
in the unconstrained system at $T_c$, it  stays finite in the 
constrained system and finite size effects are negligible even at $T_c$ 
since the constrained system is not at a critical point there. Therefore
$K$ can be obtained very precisely -- the constrained system was 
equilibrated using $2\times10^{3}$ Monte Carlo Steps (MCS) per site and a 
further averaging over $3\times10^3$ MCS was sufficient to obtain high 
quality data. Fig. 1 shows that indeed the harmonic theory result for 
$K^{-1}$($= T$) is poor near $T_c$.

Next we wish to estimate $y(l=0)$ from simulations as accurately as possible.
This was done in two ways: (i) the concentration $n_v$ of vortex pair 
excitations was measured in the unconstrained simulations of Eq.(1), (ii) the 
rejection rate $p$ of Monte Carlo moves that were rejected was measured in 
the constrained simulations (Fig. 2.). The chemical potential $\mu$ of the 
vortices could be obtained from the slope of $-\ln n_v$ or $-\ln p$ as 
a function of $T^{-1}$. We see that again our data for $p$ in the constrained 
simulations were of much superior quality because of the absence of a 
phase transition. Our estimate for $y(l=0) = \exp(-\mu/ T)$ can now be used 
together with our value for $K^{-1}(l=0)$($\simeq T + T^2/2$ from numerical 
fits to the data) in the recursion relations Eq.(5) to obtain the 
renormalized rigidity modulus $K_R$ and hence $T_c$.

The recursion relations Eq.(5) are solved numerically to obtain the 
renormalized modulus and fugacity. Using only the linearized equations (which 
can be solved analytically)  we obtain $T_c = 0.925$ which is considerably 
closer to the experimental value than the KT estimate of $1.35$ but there 
is still a significant discrepancy. However, this discrepancy vanishes when 
the full equations incorporating leading order correction terms (which 
do not affect universal behaviour and hence usually omitted) are used.
Taking leading order correction terms into account, we obtain $T_c = 0.899
\pm 0.005$ in excellent agreement with the brute force simulations! Fig.3 
presents the resulting flow diagram, which displays the importance of these
correction terms directly. 

In addition, our results offer a simple way of calculating the nonuniversal
critical amplitude $b$ (note that Fig. 3 implies that Eq.(4) holds, of course).
Critical amplitudes are usually 
notoriously difficult quantities to estimate directly from simulations.
Following Ref.\cite{amit} we define $y_0$ as the intercept
of the flows for $T > T_c$ with the $y'-$axis (see Fig. 3). The leading order 
behaviour of the correlation length $\xi \sim \pi/y_0$. Using the fact that 
$y_0^2$ has an expansion $y_0^2 = \sum_i a_i t^{i}$ with 
$t = (T - T_c)/T_c$ (see Fig. 4), we get $b = \pi/\sqrt a_1 
= 1.534 \pm .002$ which is in excellent agreement with the estimate 
$b = 1.585(9)$ obtained by Olsson\cite{OL} by directly fitting the behaviour 
of the dielectric function $\epsilon(t) = K/K_R$. 
\vskip 1cm 

\noindent
\underline{{\bf Discussion and conclusions}} Our analysis demonstrates that
the planar rotor model is fully consistent with the KT theory, but for a 
quantitatively accurate description of the transition (especially for 
nonuniversal quantities) it is indispensible that the higher order terms in 
the recursion relations Eq.(5) are taken into account. This finding implies 
that far away from $T_c$ significant corrections to Eq.(3) are expected ---
this offers an explanation why the direct simulations \cite{patra,
TC,HC,GB,JN,OL,Janke}
have difficulties in extracting the correct critical behaviour unambiguously.
In contrast, our method yields critical properties with comparatively modest 
computational effort. We expect that analogous methods can be applied to 
other models that are expected to show KT transitions; in fact, we are 
currently undertaking an extension of our approach to the controversial issue
of two- dimensional melting.
\vskip 1cm 

\noindent
\underline{{\bf Acknowledgement}} We thank Erik Luijten, C. Dasgupta
and Wolfgang Janke for 
many interesting 
discussions. One of us (S.S.) is grateful to the Alexander von Humboldt 
foundation for granting him a fellowship and P.N. thanks the 
SFB 513 for support.

\newpage
\noindent
\underline{FIGURE CAPTIONS}
\vskip 1cm 

\noindent
{\bf Fig~1.}~The inverse spinwave stiffness 
 $K^{-1}(T)$ plotted vs. temperature, for two types of simulations: (i)
The unconstrained system of $L \times L$ lattices with $L = 100$
($\Diamond$); (ii) The constrained (vortex free)
system for $L = 30 (\times), 60 (+)$ and $100(\Box)$, respectively. 
The dashed straight line represents the
result of harmonic theory ($ K^{-1} = T$) while the full curve is a fit to 
the form $T + a T^2 $ with $a = .50 \pm .01$. 
\vskip 1cm 

\noindent
{\bf Fig.~2.}~Plot of  $-\ln(n_v)$ and $-\ln(p)$ vs. the inverse temperature.
The vortex concentration $n_v$ ($\Diamond$) is calculated 
in  the unconstrained simulation of Eq.(1) for $L \times L$ lattices 
($L = 100$). The rejection ratio $p$ 
in the constrained simulation was calculated for $L = 60 (+)$ and $L = 100 
(\Box)$ respectively. The dotted line is a fit to the latter data
(near the transition temperature) yielding $2 \mu = c = 6.55 \pm .03$.
\vskip 1cm 

\noindent
{\bf Fig.~3.}~Flows of $x = (2 - \pi K)$ and $y = \exp(-\mu/T)$ under 
the action of the renormalisation group starting from a set of initial 
conditions ($\circ$) obtained from our simulations of the XY model. The
dotted lines ($y = \pm x$) are the separatrix for the linearized flow 
equations valid for flows near the fixed point $x = 0, y = 0$,
the thick lines are the actual separatrix for the nonlinear equations 
Eq. (5). Note that these curves separate flows that terminate on the critical 
line $x < 0, y = 0$ (ordered phase) from flows towards $y \rightarrow \infty$ 
(disordered phase). Arrows give the direction of the flow. 
\vskip 1cm 

\noindent
{\bf Fig.~4.}~Square of the intercept $y_0$, of the flows (Fig. 3.) with the 
$y'-$axis as a function of $t = (T-T_c)/T_c$ for $T > T_c$. The data points 
($\Diamond$) are fitted to a straight line $y' = a_1 t$. The critical amplitude 
$b$ for the correlation length $\xi \sim \exp(b t^{-1/2})$, is given by 
$b = \pi/\sqrt{a_1} = 1.534 \pm .002$.

\newpage
\noindent
\underline{FIGURES}
\vskip 1cm 

\centerline{
\epsfysize 10cm \epsfbox{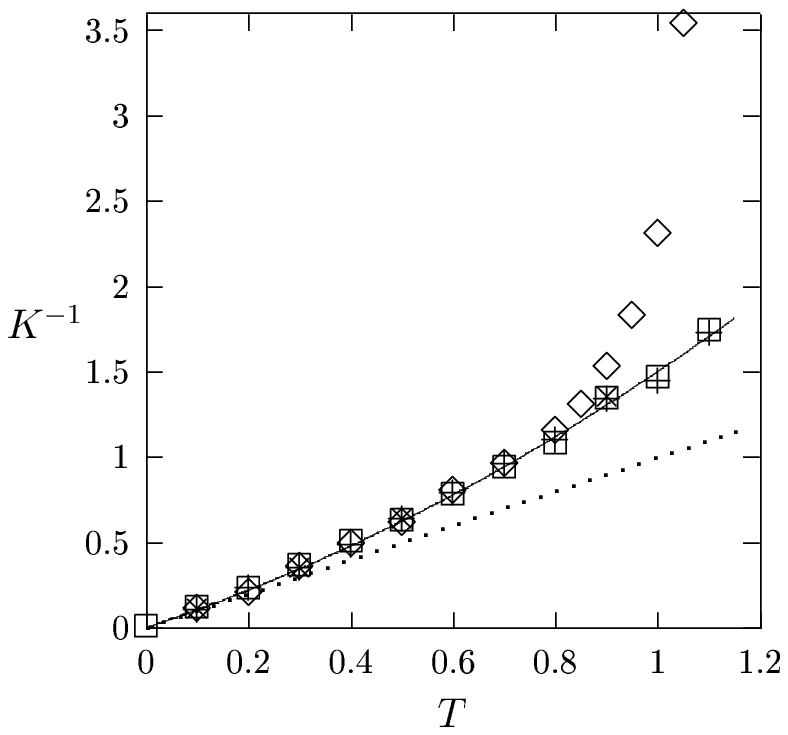}}
\vskip 3cm
\noindent
{\bf Figure 1:} \\
(Sengupta, Nielaba , Binder; Euro. Phys. Lett.)

\centerline{
\epsfysize 10cm \epsfbox{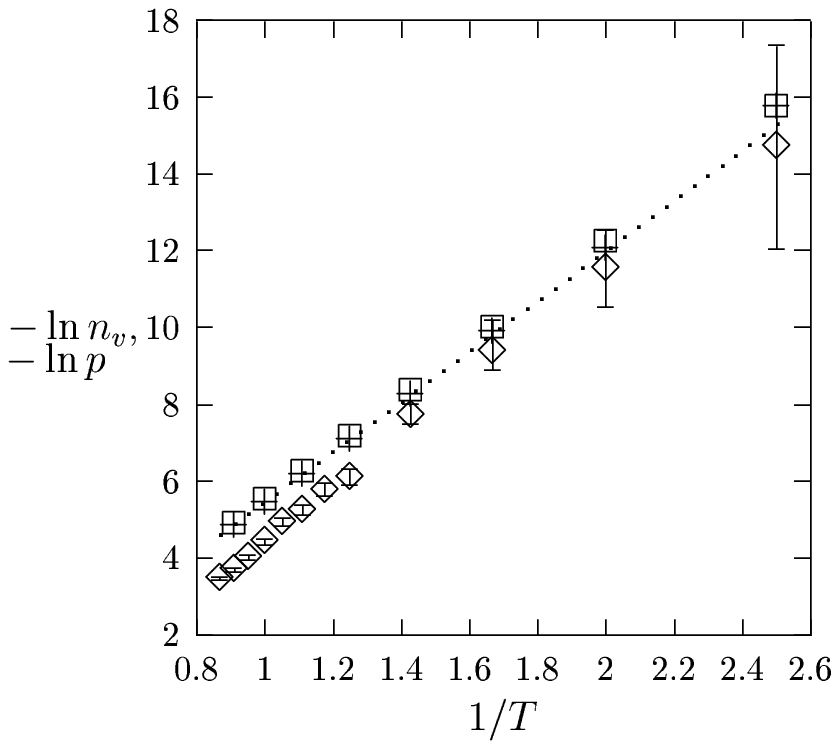}}
\vskip 3cm
\noindent
{\bf Figure 2:} \\
(Sengupta, Nielaba , Binder; Euro. Phys. Lett.)

\centerline{
\epsfysize 10cm \epsfbox{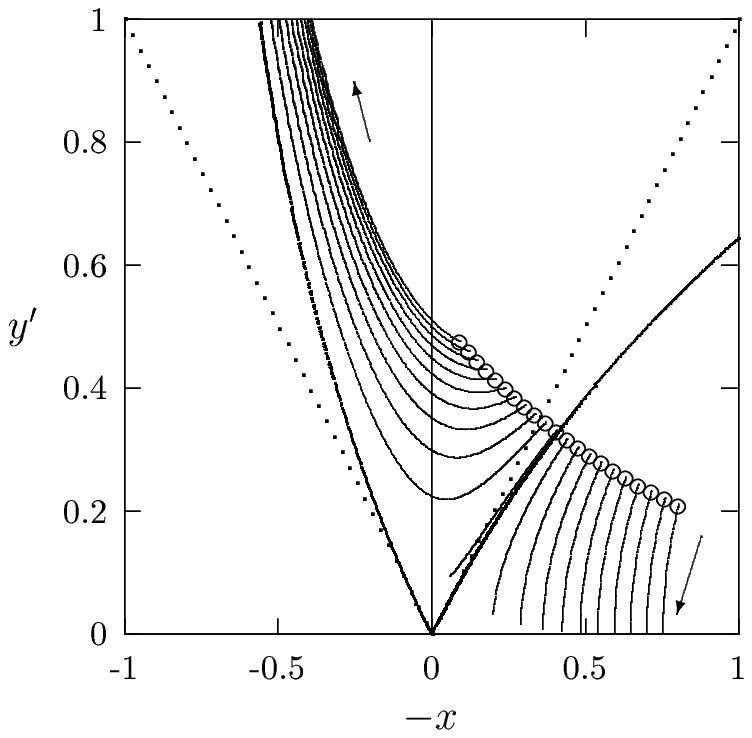}}
\noindent
{\bf Figure 3:} \\
(Sengupta, Nielaba , Binder; Euro. Phys. Lett.)

\centerline{
\epsfysize 10cm \epsfbox{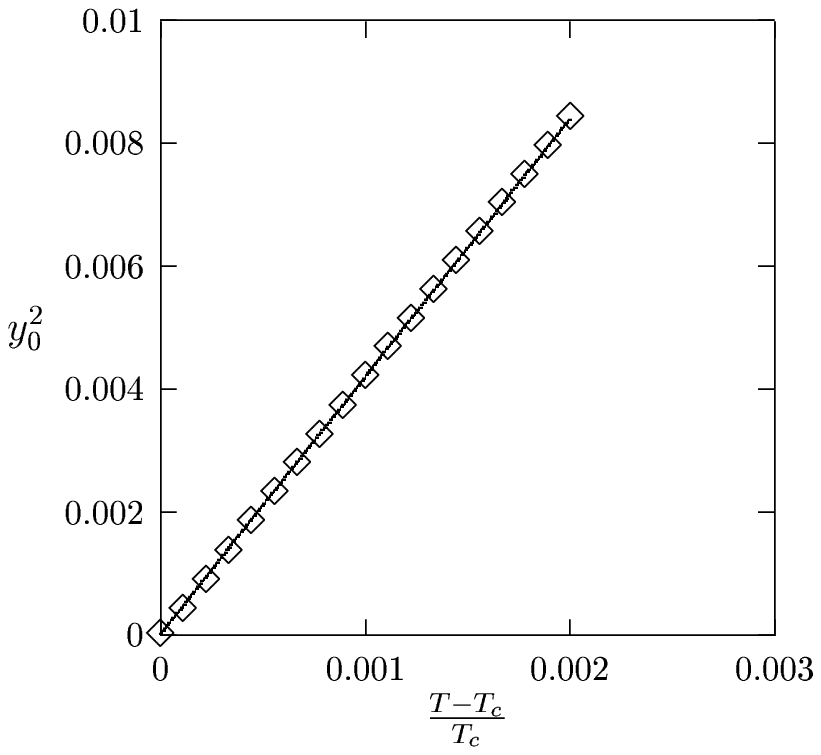}}
\noindent
{\bf Figure 4:} \\
(Sengupta, Nielaba , Binder; Euro. Phys. Lett.)

\end{document}